\documentclass[aps,pra,twocolumn,notitlepage,floatfix,tightenlines,superscriptaddress,longbibliography]{revtex4-1}

\usepackage{amssymb}
\usepackage{bbm}
\usepackage{amsmath}
\usepackage{graphicx}
\usepackage{tabularx}
\usepackage[caption=false]{subfig}
\usepackage[colorlinks=true,linkcolor=blue,anchorcolor=red,citecolor=blue,urlcolor=blue]{hyperref}

\makeatletter

\usepackage[colorlinks=true,citecolor=blue,linkcolor=blue,urlcolor=blue]{hyperref}

\makeatother

\renewcommand{\vec}[1]{\mathbf{#1}}

\newcommand{\Psivar}{\Psi_{\mathrm{var}}}

\newcommand{\veck}{\vec{k}}

\newcommand{\Cs}{$^{133}$Cs}

\newcommand{\asbg}{a_{\mathrm{bg}}}
\newcommand{\ambg}{a_{\mathrm{mm,bg}}}

\newcommand{\sumprime}{\mathop{\smash{\sideset{}{'}\sum}\vphantom{\sum}}}
\newcommand{\prodprime}{\mathop{\smash{\sideset{}{'}\prod}\vphantom{\prod}}}

\begin{document}

\title{Stability and Dynamics of Atom-Molecule Superfluids Near a Narrow Feshbach Resonance}
\author{Zhiqiang Wang}
\affiliation{Department of Physics and James Franck Institute, University of Chicago, Chicago, IL 60637, USA}
\affiliation{Hefei National Research Center for Physical Sciences at the Microscale and School of Physical Sciences, University of Science and Technology of China,  Hefei, Anhui 230026, China}
\affiliation{Shanghai Research Center for Quantum Science and CAS Center for Excellence in Quantum Information and Quantum Physics, University of  Science and Technology of China, Shanghai 201315, China}
\affiliation{Hefei National Laboratory, University of  Science and Technology of China, Hefei 230088, China}
\author{Ke Wang}
\affiliation{Department of Physics and James Franck Institute, University of Chicago, Chicago, IL 60637, USA}
\author{Zhendong Zhang}
\affiliation{E. L. Ginzton Laboratory and Department of Applied Physics, Stanford University, Stanford, CA 94305, USA}
\author{Shu Nagata}
\affiliation{Department of Physics and James Franck Institute, University of Chicago, Chicago, IL 60637, USA}
\affiliation{Enrico Fermi Institute, University of Chicago, Chicago, IL 60637, USA}
\author{Cheng Chin}
\affiliation{Department of Physics and James Franck Institute, University of Chicago, Chicago, IL 60637, USA}
\affiliation{Enrico Fermi Institute, University of Chicago, Chicago, IL 60637, USA}
\author{K. Levin}
\affiliation{Department of Physics and James Franck Institute, University of Chicago, Chicago, IL 60637, USA}
\date{\today}

\begin{abstract}
The recent observations of a stable molecular condensate emerging from a condensate of bosonic atoms
and related ``super-chemical" dynamics have raised an intriguing set of questions.
Here we provide a microscopic understanding of this unexpected stability and dynamics in
atom-molecule superfluids; we show
one essential element behind these phenomena
is an extremely narrow Feshbach resonance in \Cs~at 19.849G.
Comparing theory and experiment
we demonstrate how this narrow resonance enables
the dynamical creation of a large closed-channel molecular fraction
superfluid, appearing in the vicinity of unitarity.
Theoretically the observed superchemistry (\textit{i.e.},
Bose enhanced reactions of atoms and molecules), is found to be assisted by the formation of Cooper-like pairs of bosonic atoms that have opposite momenta.
Importantly, this narrow resonance opens the possibility to
explore the quantum critical point of a molecular Bose superfluid and related phenomena which would not be possible near a more typically broad Feshbach resonance.
\end{abstract}

\maketitle

\section{Introduction}

Pairing in ultracold quantum gases and the preparation of quantum degenerate molecules
have been long sought-after goals~\cite{Bohn2017,Carr2009,Quemener2012}
for some time in the cold atom community.
It provides access to new forms of many body
physics and quantum metrology.
Historically, experiments in pursuit of such quantum degenerate ultracold molecules often have been
hindered by cooling challenges and collisional loss~\cite{Langen2023,Carr2009,Bohn2017}.
That said, there have been successes, more numerous for fermionic systems~\cite{Langen2023,DeMarco2019,Jochim2003}. Recently
a stable ~\Cs$_2$
molecular condensate consisting of bosonic \Cs~atoms has been reported~\cite{Zhang2021,Zhang2023}. Here pairing interactions were induced in an atomic condensate
based on a
$g$-wave Feshbach resonance at 
$B_0 =19.849(2)$G~\cite{Zhang2023}.

In this Letter 
we show that essential for observing this molecular
superfluid phase and a dynamically generated superchemistry~\cite{Zhang2023}
is a narrow Feshbach resonance used in the experiment to generate molecules~\cite{Chin2010}. This, in \Cs, has a width
$\Delta B = 8.3(5)$ mG~\cite{Zhang2023}, which is
three to four orders of magnitude smaller than typically considered
in $^{7}$Li~\cite{Pollack2009}, $^{85}$Rb~\cite{Donley2002,Holland2001} and $^{39}$K \cite{Eigen2018,DErrico2007}.
See Table~\ref{tab:CsRbK}.
[Here we use the
dimensionless resonance width parameter from the so-called many-body classification scheme~\cite{Ho2012}.] 
This narrow resonance provides an explanation for the much wider
stability regime and, importantly, enables access to the atom-molecule
quantum critical point (QCP)~\cite{Radzihovsky2004,Romans2004}
near the Feshbach resonance.

Through a comparison between theory and experiment, we demonstrate
how a magnetic field quench, which sweeps an atomic superfluid to near unitarity, 
leads to a superfluid having a large closed-channel molecular fraction.
Our theoretical analysis identifies an important role for out-of-equilibrium,
non-condensed Cooper-like pairs, which are created by the Feshbach coupling during 
a transient stage. 
These are necessarily distinct from
quantum depletion effects~\cite{StamperKurn1999} which arise due to repulsive background scattering.
We find the Feshbach-coupling induced pairs fully
participate in the coherent oscillations of the condensates that follow.
That the associated oscillation frequency scales with the number of atoms reflects a coherent quantum chemical process stimulated by Bose-ehanchement, \textit{i. e.} superchemistry~\cite{Heinzen2000,Vardi2001,Richter2015}.

\begin{table*}[htp]
    \center
    \caption{Experimental parameters for Feshbach resonances in different Bose gases. 
    In this table, $m_1$ is the atomic mass, $B_0$ is the experimental resonance point, 
    $\Delta \mu_m$ is the magnetic moment difference between a pair of atoms in the open channel and a molecule in the closed channel, 
    $\Delta B$ is the resonance-width in magnetic field, $\asbg$ is the atom-atom background scattering length, and $n$ is the experimental number density of atoms. 
    In the last column, $x \equiv (k_n r_*)^{-1}= |k_n \asbg|^{-1} |\Delta \mu_m \Delta B| / E_{\mathrm{bg}}$ is the dimensionless resonance-width parameter introduced in Ref.~\cite{Ho2012}. 
    Here, $k_n=(6\pi^2 n)^{1/3}$ and $E_{\mathrm{bg}}\equiv \hbar^2/(m_1 \asbg^2)$. 
    The data for $^{133}$Cs, $^{85}$Rb, and $^{39}$K are collected from Refs.~\cite{Zhang2021,Zhang2023}, Refs.~\cite{Donley2002,Holland2001a}, and Refs.~\cite{Eigen2018,DErrico2007}, respectively. $a_B$ is the Bohr radius, and $\mu_B$ is the Bohr magneton. 
     }
    \begin{tabular}{  c   c   c  c  c  c c c }
       \hline \hline
               Atom                    &  ~~~ $m_1$ (a. m. u.)~~~      &   ~~~~ $B_0$  ~~~~   &    ~~~~ $\Delta \mu_m$  ~~~~    &    ~~~~ $\Delta B$ ~~~~     &     ~~~~ $\asbg$  ~~~~    &  ~~~~ $n$ (cm$^{-3}$)  ~~~~                             &       $x=(k_n r_*)^{-1}$    \\
        \hline
        ~~ $^{133}$Cs  ~~     &    ~~  132.91~~    &     ~~ 19.849 G ~~       &   ~~   0.57  $\mu_B$  ~~         &    ~~  8.3 mG  ~~                &     ~~   163 $a_B$ ~~            &    ~~  $2.9 \times 10^{13}$  ~~  &  $ 10^{-1}$ \\
         ~~ $^{85}$Rb  ~~     &     ~~  84.91 ~~   &        ~~ 155 G ~~       &      ~~$ - 2.23$ $\mu_B$ ~~~       &    ~~ 11.06 G ~~~             &     ~~ $- 450$ $a_B$   ~~                                    &     ~~  $3.9\times 10^{12}$  ~~   &   $10^3$         \\
         ~~ $^{39}$K  ~~     &       ~~  38.96~~  &     ~~   402.7 G  ~~    &    ~~~  1.5  $\mu_B$ ~~~       &         ~~~     52 G   ~~~        &        ~~  $- 29$  $a_B$ ~~         &   ~~  $5.1 \times 10^{12}$ ~~ &   $ 10^2$                              \\
             \hline \hline
          \end{tabular}
    \label{tab:CsRbK}
\end{table*}

The theoretical framework we employ incorporates a narrow Feshbach resonance and provides an
integrated description of both the equilibrated system
and the non-equilibrium
dynamics.
This narrow resonance ensures that
the molecules near unitarity are predominantly closed-channel like,
in contrast to the open-channel dominated bound states studied previously \cite{Donley2002,Claussen2002,Makotyn2014,Eigen2017,Eigen2018}. 
The narrowness of the resonance, (combined with a repulsive inter-molecular interaction~\cite{Zhang2021}), leads to the unexpected stability at equilibrium~\cite{Koetsier2009,Jeon2002,Mueller2000,Basu2008} . 

\begin{figure}[tp] 
\includegraphics[width=3.5in,clip]
{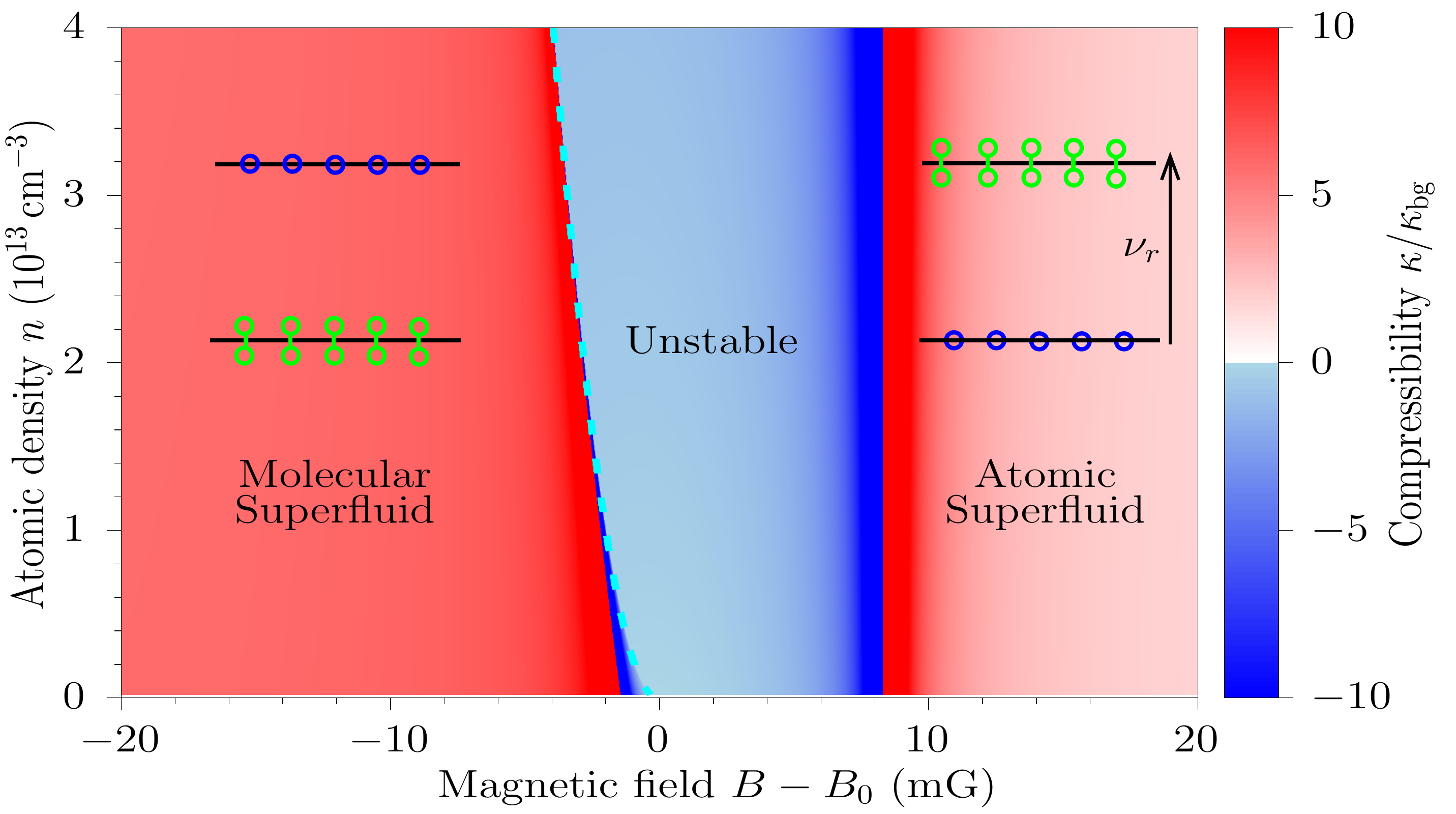}
\caption{Ground state stability phases for the $g$-wave
resonance of \Cs~at $B_0 = 19.849$ G with width $\Delta B = 8.3$ mG.
Plotted is a map of the compressibility  $\kappa= \partial n/\partial \mu$ as a function of atomic density $n$ and magnetic field $B$, measured relative to $B_0$.
$\kappa$ is normalized by $\kappa_{\mathrm{bg}}=m_1/(4\pi \hbar^2 a_{\mathrm{bg}})$ with $m_1$ the atomic mass and and $a_{\mathrm{bg}}$ the background scattering length. The
atomic superfluid and molecular superfluid phases 
are stable in the red region, unstable in the blue.
Indicated are the energy levels of atoms  (blue circles) and molecules
(green pairs) which characterize the phases with the molecular energy $\nu_r$ that is approximately $\propto (B-B_0)$~\cite{Lange2009}. 
The dashed line (cyan) is the expected QCP, obtained without taking into account the stability issue.
 }
\label{fig:1}
\end{figure}

To address this stability we turn first to the theoretically calculated phase diagram depicted in
Fig.~\ref{fig:1}. The figure shows that there are
two superfluid phases: the
atomic superfluid (ASF) in which both atomic
and molecular condensates co-exist and the molecular superfluid (MSF)
where the atomic condensate is missing~\cite{Radzihovsky2004,Romans2004}.
We will return to this figure in more detail later, but note a central conclusion: that
there is only a narrow range of magnetic fields,
mostly associated with the region between the so-called QCP
and the zero crossing of the atomic scattering length, where instability is present.

\subsection*{Experimental background}

Our experiments start with a Cs BEC of 23,000 atoms at 22 nK in a pancake-like harmonic trap. The trap frequencies are ($\omega_x$, $\omega_y$, $\omega_z$) = $2\pi\times$(24, 13, 74) Hz. To initiate the non-equilibrium dynamics in the atomic and molecular channels, we quench the magnetic field to near the $g-$wave Feshbach resonance.
After a variable evolution time, we decouple the atomic and molecular channels by quickly switching the magnetic field far below the resonance so that we
can independently detect the population and temperature in each channel by focused time-of-flight (TOF) imaging. 
In this imaging molecules are first released into an isotropic harmonic trap for a quarter trap period before being dissociated above the Feshbach resonance.
We also study molecule dissociation dynamics. For these latter experiments we first prepare a molecular condensate with a 23\% BEC fraction~\cite{Zhang2021,Supplement}. Then the magnetic field is quenched close to the resonance and we monitor the atom number resulting from dissociation.
For more details about the TOF imaging and experimental timeline, see Appendix~\ref{app:experiment}.

\section{Theoretical framework and Results}

The narrowness of the resonance requires us to consider a 
theoretical framework associated with ``two-channel" physics, in contrast
to effective one-channel descriptions
~\cite{Natu2013,Rancon2013,Kain2014,Sykes2014,Corson2015,Menegoz2015,Yin2016,VanRegemortel2018,MunozdelasHeras2019}.
The Hamiltonian 
$\hat{H} = \hat{H_1} + \hat{H_2} + \hat{H_3}$,
contains 
a kinetic energy ($\hat{H_1}$) for the two
species (open channel atoms and closed channel molecules), the intra-species repulsive interactions $g_\sigma$~\cite{Zhang2021}, and
the Feshbach coupling $\alpha$.
Here,
\begin{subequations}  \label{eq:Hamiltonian}
\begin{align}
 \hat{H_1} & = \sum_{\veck}\sum_{\sigma=1}^2  h_{\sigma \veck}  \, a_{\sigma \veck}^\dagger a_{ \sigma \veck}, \\
\hat{H_2} &=
 \frac{1}{V}\sum_{\veck_i} \sum_{\sigma=1}^2 \frac{g_\sigma}{2} a^\dagger_{ \sigma \veck_1 } a^\dagger_{ \sigma \veck_2 } a_{  \sigma, \veck_3} a_{ \sigma, \veck_1 +\veck_2 -\veck_3 }, \label{eq:H2}
\\
\hat{H_3}
&= -\frac{\alpha}{\sqrt{V}}   \sum_{\veck_i}  \big( a^\dagger_{1 \veck_1 } a^\dagger_{1 \veck_2 } a_{ 2 ,\veck_1+\veck_2} + h. c.\big).   \label{eq:H3}
\end{align}
\end{subequations}
The subscripts $\sigma={1}$ and ${2}$ represent open channel atoms and closed channel molecules, respectively.
$V$ is the volume, and $V^{-1}\sum_{\veck} = \int^\Lambda d \veck/(2\pi)^3$ where $\Lambda$ is a cutoff, needed to regularize an ultraviolet divergence.
We assume three-dimensional isotropy and ignore trap effects in our theory, 
as they do not affect qualitative conclusions.

In  $\hat{H_1}$, $h_{1 \veck}=(\hbar\veck)^2/2 m_1 -\mu$ and $h_{2 \veck}=(\hbar \veck)^2/2 m_2 -(2\mu -\nu)$ with $m_2=2 m_1$, $\mu$
the chemical potential, and $\nu$ the bare-molecule state detuning. 
We distinguish
$\nu$ from the detuning $\bar{\nu} \equiv \Delta \mu_m (B-B_0)$ through
a $B$-independent constant;
here, $\Delta \mu_m>0$ is the difference in magnetic moments of the two
channels, and $B_0$ is where the atomic two-body scattering length diverges, as in experiment.
The eigenenergy of dressed molecules, denoted as $\nu_r$ in Fig.~\ref{fig:1}, is 
nearly equal to $\bar{\nu}$ for the g-wave resonance of \Cs~at $B_0=19.849$G, except when $|B-B_0| \ll 1$mG~\cite{Chin2010,Lange2009}. 
The $\alpha$ in $\hat{H_3}$, given by $\alpha = \sqrt{ (2\pi \hbar^2
a_{\mathrm{bg}} /m_1) \Delta \mu_m \Delta B}/ \big[ 1-  (2/\pi)
a_{\mathrm{bg}} \Lambda \big]$(see details in Appendix~\ref{app:regularization}), is
chosen such that it reproduces the experimental resonance width $\Delta B$
in the two-body scattering limit.

\begin{table}[tp]
    \center
    \caption{Parameters used in the numerical simulation for $^{133}$Cs. In this table, $k_n=(6\pi^2 n)^{1/3}$ and $E_n=\hbar^2 k_n^2/2 m_1$ with $n=2.9\times 10^{13}$cm$^{-3}$.  }
    \begin{tabular}{  c   c  c  c   }
        \hline \hline
        ~~ $\Lambda$~~               &     ~~ $\alpha$ ~~                       &    ~~ $g_1$ ~~                                &     ~~ $g_2$  ~~             \\
       \hline
            ~~ $\pi$ $k_n$ ~~        &    ~~ 1.6 $E_n/k_n^{3/2}$ ~~  &    ~~ 3.15 $E_n/k_n^3$ ~~              &     ~~  2.30 $E_n/k_n^3$  ~~          \\
        \hline \hline
          \end{tabular}
    \label{tab:parameters}
\end{table}

\begin{figure*}
\includegraphics[width=6.5in,clip]
{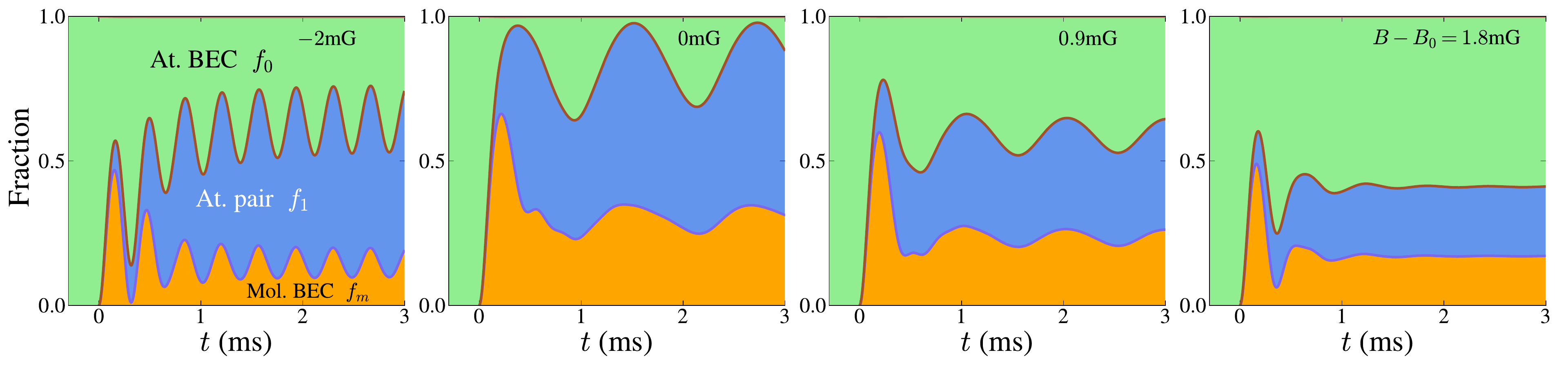}
\caption{Calculated coherent atom-molecule dynamics near resonance, obtained after a quench 
of a pure atomic condensate (at $t=0$) 
to the four indicated magnetic fields $B-B_0$. 
Shown are atomic condensate fraction $f_0=|\Psi_{10}|^2/n$ (green), atomic pair fraction $f_1=n_1/n$ (blue), and molecular condensate fraction $f_m=2|\Psi_{20}|^2/n$ (orange) versus
time $t$.
Non-condensed molecules are negligible.
Here the total particle density is set to the experimental value of $n= 2.9 \times 10^{13}$cm$^{-3}$~\cite{Supplement}.
}
\label{fig:4}
\end{figure*}

To address \textit{both} the statics and dynamics in a unified manner, we adopt a variational wavefunction,
$\vert \Psivar (t) \rangle   =$
\begin{align}
 \frac{1}{\mathcal{N}(t)}  e^{ \sum_{\sigma=1}^2 \Psi_{\sigma 0}(t) \sqrt{V}  a^\dagger_{\sigma 0}   
 + \sum_{\veck}^\prime \sum_{\sigma=1}^2  \chi_{\sigma \veck}(t) \; a^\dagger_{\sigma \veck} a^\dagger_{\sigma -\veck} } \vert 0 \rangle.    \label{eq:Psivart}
\nonumber
\end{align}
In the exponent the $\veck-$sum is over half of $\veck-$space.
$\Psi_{\sigma 0}$ and $\chi_{\veck}$ are complex variational parameters, which are time-dependent (-independent) for our study of dynamics (statics). $|0\rangle$ is the vacuum that is annihilated by all $a_{\sigma\veck}$. $\mathcal{N}(t)$ is the normalization factor.
Here, in the spirit of generalized Bogoliubov theory,  only pair-wise correlations are included in the exponent of the variational wavefunction, which can be generallly
justified by the experimental observation~\cite{Zhang2023}
of undamped coherent oscillations of the populations which persist to long times.

The many-body dynamics associated with $\hat{H}$ can be approximated through the variables $\Psi_{\sigma 0}(t)$
and $\chi_{\sigma \veck}(t)$,  which in turn are derived from the action~\cite{Haegeman2011,Shi2018,Kramer2008}
$\mathcal{S}[\Psi_{\sigma 0}^*(t), \Psi_{\sigma 0}(t), \chi_{\sigma \veck}^*(t),  \chi_{\sigma \veck}(t)]=  \int d t  \langle \Psivar(t) | (i\hbar) \partial_t \Psivar(t) \rangle -  \langle \Psivar(t) \vert \hat{H} \vert \Psivar(t) \rangle$. 
Minimizing $\mathcal{S}$ with respect to $\{\Psi_{\sigma 0}^*, \Psi_{\sigma 0}, \chi_{\sigma \veck}^*,  \chi_{\sigma \veck} \}$ leads to the following dynamical equations~\cite{Supplement},
\begin{subequations}  \label{eq:dynamics}
\begin{align}
i \hbar \frac{d}{dt} \Psi_{\sigma 0} & = (  h_{\sigma \veck=0}  + g_\sigma |\Psi_{\sigma 0}|^2  + 2 g_\sigma n_\sigma ) \Psi_{\sigma 0}  
+ g_\sigma  \Psi_{\sigma 0}^* x_\sigma   \nonumber \\ 
&  - \delta_{\sigma, 2} \, \alpha  ( x_1 + \Psi_{10}^2 ) - \delta_{\sigma,1} \, 2 \alpha \Psi_{10}^* \Psi_{20}, \\
i \hbar \frac{d}{dt} x_{\sigma \veck} & =  2  \big[  h_{\sigma \veck} + 2 g_\sigma ( |\Psi_{\sigma 0}|^2 + n_\sigma )   \big] x_{\sigma \veck}  +  \nonumber \\
&   \big[g_\sigma  (x_\sigma + \Psi_{\sigma 0}^2)  - \delta_{\sigma, 1} \, 2 \alpha \Psi_{20} \big] (2 n_{\sigma \veck}+1),
\end{align}
\end{subequations}
where $\delta_{\sigma,\sigma^\prime}$, with $\{\sigma,\sigma^\prime\}=\{1,2\}$, is the Kronecker delta. 
We relegate detailed derivations of these equations to Appendix~\ref{app:Eq4derivation}.
In the above $\Psi_{\sigma 0}  \equiv \langle a_{\sigma 0} \rangle / \sqrt{V}$,
the ``Cooper pair"- like correlation~\cite{Holland2001a} 
$ x_{\sigma \veck}   \equiv  \langle a_{\sigma \veck}  a_{\sigma -\veck} \rangle = \chi_{\sigma \veck} / ( 1-|\chi_{\sigma \veck}|^2)$,
$ n_{\sigma \veck}  \equiv \langle a_{\sigma \veck}^\dagger a_{\sigma \veck} \rangle = |\chi_{\sigma \veck}|^2 / (1-|\chi_{\sigma \veck}|^2)$,
$ x _\sigma   =  V^{-1} \sum_{\veck \ne 0} x_{\sigma \veck }$,  
and $ n_\sigma  =  V^{-1} \sum_{\veck \ne 0} n_{\sigma \veck}$.
Here, $\langle \cdots \rangle \equiv \langle \Psivar | \cdots | \Psivar \rangle$. 
$x_{\sigma \veck}$ is the expectation value of the (Cooper-like) pairing field for atoms or molecules.
Note that both $x_{\sigma \veck}$ and $n_{\sigma \veck}$ are not independent. 
To obtain the dynamics we solve Eq.~\eqref{eq:dynamics} together with the constraint:
$ n = (|\Psi_{10}|^2 + n_1) + 2 (|\Psi_{20}|^2 + n_2)$.

An advantage of working with the variational scheme is that the statics at equilibrium can be addressed simultaneously with the dynamics.  
At equilibrium, one minimizes the trial ground state energy $\langle \hat{H} \rangle$ instead of $\mathcal{S}$ with respect to the same set of variational variables, leading to a set of self-consistent conditions that
are nearly identical to Eq.~\eqref{eq:dynamics} except that the time derivatives in the latter are set to zero.

For all figures presented here we use parameters for \Cs~based on 
Refs.~\cite{Zhang2021,Zhang2023}, (provided in Tables~\ref{tab:CsRbK} and \ref{tab:parameters}),
which have been chosen to reproduce the experimental resonance width $\Delta B$ and the atom-atom background scattering length $a_{\mathrm{bg}}$.
Knowing the density $n$, $k_n \equiv (6\pi^2 n)^{1/3}$, and $E_n =\hbar^2 k_n^2/(2m_1)$
we can calibrate the units of time in our dynamical
calculations in terms of milli-seconds (ms) and thus compare theory directly with experiment.

Solving the static version of Eq.~\eqref{eq:dynamics} together with the number density constraint,
 we obtain the equilibrium values of $\Psi_{\sigma 0}, x_{\sigma}, n_{\sigma}, \mu$, etc.
as a function of both the detuning $\bar{\nu}$ and the total density $n$. 
To establish stability in the two channel system we numerically
compute the compressibility $\kappa=\partial n/ \partial \mu$.  

Depending on whether $\kappa$ is positive (stable) or negative (unstable) the phase diagram in Fig.~\ref{fig:1} can be divided into
three regimes: stable MSF phase, stable ASF phase, and an unstable regime near $\bar{\nu}=0$. 
Stability in the MSF phase 
depends on an inter-molecule repulsion, $g_2>0$.
The stable MSF phase can persist to a regime well within the resonance width $\Delta B$ 
around unitarity,
and just to the left of the presumed QCP.
This is a unique and important characteristic of a narrow resonance as compared with a wide
resonance. In the latter case the MSF-Unstable phase boundary in Fig.~\ref{fig:1} is pushed to the far left and well separated from the QCP~\cite{Wang2023}.

We turn next to our theoretical results for quenched dynamics, obtained from Eq.~\eqref{eq:dynamics}. 
We start with a pure atomic condensate, abruptly change the detuning to final values on either the positive or negative side of resonance,
and then monitor the subsequent dynamical evolution of each component.
The results presented in Fig.~\ref{fig:4} show, after the quench, how the initially large
atomic condensate contribution is quickly converted into a closed-channel molecular
condensate (orange) as well as non-condensed pairs (blue).
The most notable features are persistent oscillations in all components,
seen for $B -B_0 \lesssim 1$ mG, which are most pronounced near unitarity. 
While the pairs and atomic condensate oscillate out of phase,
 the molecular condensate and pairs are nearly in phase.
The pairs lead to very little dephasing on the molecular side,
but above resonance for $B -B_0 \gtrsim 1$ mG the oscillations are completely damped. 

The calculation shows a substantial generation of atom pairs and
molecules only within a narrow range of the Feshbach resonance.
This can be understood as deriving from many-body entanglement generated in the dynamics
as near the resonance the system is most strongly
correlated. A quench can, thus, spread this entanglement over a larger portion of Hilbert space, 
thereby generating more correlated pairs and molecules near unitarity. 

Also notable is an asymmetry (see also Appendix~\ref{app:pair}) in the pair production between negative and positive detunings which
can be understood using the energy level diagram of Fig.~\ref{fig:1}. 
For sweeps to the molecular side of resonance,
energy conservation requires that the energy loss in a conversion from an atomic to
molecular condensate
be compensated by making 
more atom-pairs
appear at higher energies.

Our results show rather good agreement with the following features
of the experimental observations in Ref.~\cite{Zhang2021,Zhang2023}.
We see a rapid relaxation toward a quasi-equilibrated phase where oscillations
persist. These oscillations have a strong density dependence 
(associated with quantum ``superchemistry") which will
be addressed in more detail below. 
As will also be evident, the bulk of the closed channel molecule production takes place over a 
relatively narrow range
of fields roughly within the resonance width of $\Delta B \sim 8$ mG.

\section{Comparison between theory and experiment}

\begin{figure*}
\includegraphics[width=6.5in]
{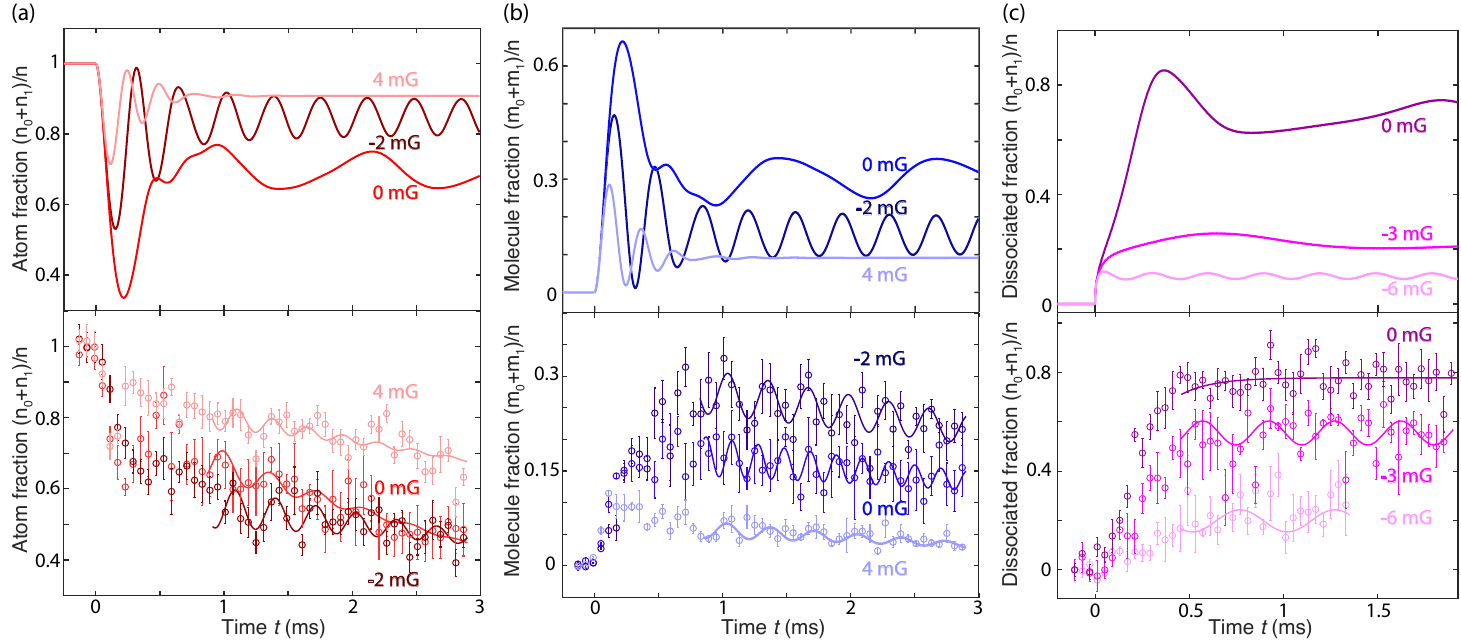}
\caption{Coherent atom-molecule dynamics, in theory (top panels) and experiment (bottom).
Curves with the same color should be compared.
Panels (a) and (b) respectively denote atomic and molecular channels, when an atomic condensate is quenched to different values of $B-B_0$ (in mG) near the resonance.
$n_0=|\Psi_{10}|^2$, $m_0=2 |\Psi_{20}|^2$, $m_1=2 n_2$. (For definitions of $\Psi_{10},\Psi_{20},n_1$ and $n_2$ see text). 
Plotted on the vertical axis
in (c) is the fraction of molecules dissociated
when a molecular condensate is quenched to different $B-B_0$ values. 
Solid lines (bottom panels) are fits to the data following the procedure in Ref.~\cite{Zhang2023}; error bars represent one standard deviation from the mean.
The particle density $n= 2.9 \times 10^{13}$cm$^{-3}$~\cite{Supplement}. 
}
\label{fig:FigAtm}
\end{figure*}

\begin{figure}[htp] 
\includegraphics[width=0.75\linewidth,clip]
{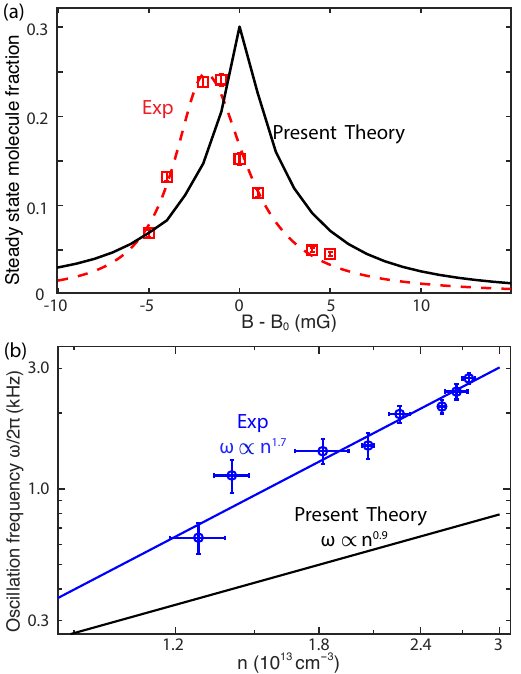}
\caption{
(a) Closed-channel molecule fraction obtained as a time-average after $t\approx 1$ms~\cite{Supplement},
for the state reached after a quench as in Figs.~\ref{fig:4} and \ref{fig:FigAtm}(a,b).
The red open squares are experimental data for $n=2.9\times 10^{13}$cm$^{-3}$; the red dashed line is a guide to the eye.
(b) Density $n$ dependence of the oscillation frequency $\omega$ near unitarity.
The blue circles, (with the blue line a power law fit), are from previous experiments~\cite{Zhang2023} compared with theory (black solid line).
}
\label{fig:3}
\end{figure}

Specific plots illustrating the atomic and closed-channel molecule populations
are presented
in Figs.~\ref{fig:FigAtm}(a) and (b) respectively with top panels for theory and bottom for
experiment.
These are to be associated with
the dynamics
after a quench of an atomic condensate to different final detunings $\bar{\nu}$. 
We note that comparing curves with the ``same" values of magnetic
field in Fig.~\ref{fig:FigAtm}, a field recalibration might be
considered as we will see in Fig.~\ref{fig:3}(a) that there is a small off-set in $B-B_0$ of  the order of
$2$mG between where the molecular fraction reaches a maximum in the theory as compared with experiment.
One can also see from Fig.~\ref{fig:FigAtm} that a more significant difference between
theory and experiment
is associated with the initial large overshoot, particularly of the molecular
contribution, 
which is absent in the experiment. This difference is likely
due to inelastic particle-loss processes, which are most prevalent in the molecular channel.
Another contributing factor to the difference is the fact that there is a non-negligible delay in transitioning the magnetic fields in the experiment,
which can partially obscure or interfere with the early time measurements where the overshoot is observed in theory.
At late times $t \gtrsim 1$ ms, the experimentally observed oscillations of atoms and molecules in Figs. ~\ref{fig:FigAtm}(a) and \ref{fig:FigAtm}(b) are not completely out of phase (see also Fig. 3a of Ref.~\cite{Zhang2023}). This hints that there exists some small loss process that is coherent and persistent.
Such a coherent loss process, which is absent in our theoretical simulations, will be investigated in future work. 

The current narrow resonance of \Cs~provides a unique opportunity to probe new issues which are not present in
the moderately wide resonances typically used~\cite{Donley2002, Claussen2002,Makotyn2014,Eigen2017,Eigen2018}. 
In particular we can consider the post-quench dynamics for systems near unitarity, which are initially prepared as a molecular superfluid state.
Theory predicts that the steady-state molecule dissociation fraction, reached after a transient stage,
will increase with the final detuning $\bar{\nu}$ as 
$|\bar{\nu}| \rightarrow 0$.
It is also interesting to note that
a residual steady-state oscillation is observed in experiments
which appears robustly in theory provided a very small atomic condensate ``seed" is introduced to the initial molecular superfluid state.
Indeed, both these observations can be verified through
a direct comparison between theory and experiment  
in Fig.~\ref{fig:FigAtm}(c)
where the agreement is quite reasonable.

We turn to another comparison in Fig.~\ref{fig:3}(a) 
which addresses~\footnote{In the experiments the fraction
plotted is for both condensed and non-condensed closed-channel molecules,
whereas in theory almost all closed-channel molecules are condensed.}
the question at what range of magnetic
fields, after a quench, is there an appreciable production of closed-channel molecules.
Fig.~\ref{fig:3} (a) plots the corresponding fraction, which is for the quasi-steady state
associated with a time where the molecular fraction saturates, as in Fig.~\ref{fig:FigAtm}(a,b).
From the figure one sees that in both theory and experiment, not only is
the fraction largest in the near-vicinity of resonance but the maximum
in both is between 20\% and 30\%. It is interesting to observe that this
maximum closed-channel molecular fraction, (which has been a topic
of interest both for dynamically generated~\cite{Mark2005} and equilibrated superfluids),
is significantly lower than found for Fermi-superfluids ~\cite{Koehler2006}.

The phenomenon of ``quantum superchemistry"~\cite{Zhang2023,Heinzen2000,Vardi2001,Richter2015}
is of particular interest to explore as it is reflected in a dependence of the oscillation frequency $\omega$ 
on the density $n$.
Such a density dependence, indicative of
a many-body Bose enhancement of chemical reactions, 
can be quantified as a power law
$\omega \propto n^\gamma$ when $B =B_0$. Experiments find that $\gamma \approx 1.7$ while in the present theory
$\gamma \approx 0.9$.
Results from both theory and experiment are shown in Fig.~\ref{fig:3}(b) (see also Appendix~\ref{app:comparison}), 
although a more systematic comparison would require the inclusion of trap effects in the theory.
Despite the fact that the exponents show some differences what is important here
is the observation of a Bose-enhanced chemistry even in the presence
of Cooper-like pair excitations at finite momenta.
While one might have expected these pairs to undermine or dissipate the oscillations,
they appear to participate fully and maintain their coherence. 

To interpret the superchemical oscillations, two phenomenological 
Hamiltonians, associated with two-body and three-body models, were used in Ref.~\cite{Zhang2023}, 
which contemplated only two modes, the atomic ($\Psi_{10}$) and molecular condensates ($\Psi_{20}$). 
We emphasize that even though the present two-channel Hamiltonian in Eq.~\eqref{eq:Hamiltonian} 
only contains Feshbach coupling and pair-wise density-density interactions,
it can induce the three-body processes discussed in Ref.~\cite{Zhang2023}. 
These arise through scattering events that are of higher order than linear in the Feshbach coupling constant. 
This should not be surprising since the Hamiltonian in Eq.~\eqref{eq:Hamiltonian} has been used in the literature~\cite{Bedaque2000}
to discuss three-body recombination and related Efimov physics.

\section{Conclusions}

In conclusion, in this paper, we have shown that for the particular narrow $g$-wave Feshbach resonance at $B_0 =19.849(2)$G in
\Cs~the ground-state phase diagram around the predicted quantum critical point is interrupted only by a narrow region
of instability. In the future one can study this QCP from
the molecular side, which is in contrast to the situation for a typical ``wide" resonance where 
this critical point
is inaccessible~\cite{Wang2023}.

We have also addressed the post-quench dynamics around this resonance, primarily focusing on the coherent oscillations introduced by the quench.
We have shown that for such an extremely narrow Feshbach resonance an appreciable
fraction of closed-channel molecules can be produced from quenching an atomic BEC.
Here, we provide comparisons between theory and experiment for the post-quench dynamics, which involves
3 constituents that participate in the quasi-steady-state oscillations: 2 condensates along with correlated pairs of atoms.
We caution that this paper is not focused on arriving at a precise quantitative agreement between theory and experiment,
as various inelastic scattering processes such as three-body loss, atom-molecule, and molecule-molecule collisional losses are not included
in the theoretical modeling.
It is well known that these loss processes are extremely challenging to address for Bose gases near unitarity, in both theory and experiment.

Our work here emphasizes that the experimentally 
observed, quench induced coherent oscillations~\cite{Zhang2023}
are consistent with the existence of non-condensed pairs, which
importantly do not
undermine the
highly collective nature of the observed superchemistry.
This follows because the
pairs participate fully along with both atom and molecule condensates in the coherent dynamics.
In the future, it will be interesting to look for more direct evidence of these pairs,
using either pair-pair correlations as in Ref.~\cite{Tenart2021} or matter-wave jet emissions as in Ref.~\cite{Clark2017}. 

We end by noting that our current studies of the $g$-wave resonance in $^{133}$Cs, which provide the first
observation of a molecular BEC consisting of bosonic atoms, suggest an important role for our
paper, as it serves to guide and encourage future efforts in other atomic gases
with narrow resonances such as $^{23}$Na~\cite{Xu2003}, $^{87}$Rb~\cite{Duerr2004}, $^{168}$Er~\cite{Frisch2015}, where many of these
same conclusions should apply.

\section*{Acknowledgments}
We thank Paul S Julienne, Qijin Chen, Zoe Yan, and  Leo Radzihovsky for helpful discussions and communications at different stages of this project. 
This work is supported by the National Science Foundation under Grant No. PHY1511696 and PHY-2103542,
and by the Air Force Office of Scientific Research under award number FA9550-21-1-0447.
Z.Z. acknowledges the Grainger Graduate Fellowship and the Bloch Postdoctoral Fellowship. 
S.N. acknowledges support from the Takenaka Scholarship Foundation.
Z. W. is supported by the Innovation Program for Quantum Science and Technology (Grant No. 2021ZD0301904). 
We also acknowledge the University of Chicago's Research Computing Center for their support of this work.

\appendix
\numberwithin{equation}{section}

\section{Preparation and Detection of Atomic and Molecular BECs}
\label{app:experiment}

\begin{figure}[h!]
    \centering
    \includegraphics[width=0.85\linewidth]{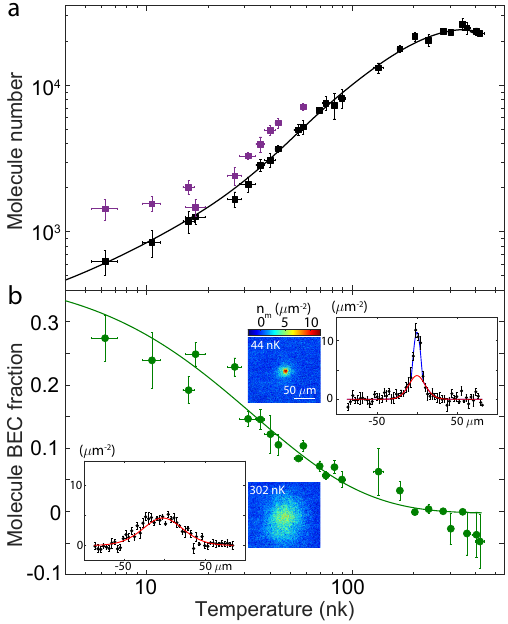}
    \caption{Molecule number and molecular BEC fraction prepared at different temperatures. \textbf{a}, Number of molecules created by associating atoms in ultracold atomic gases at different temperatures. The black (purple) data points are from 16.7 ms focused time-of-flight (ToF) (\textit{in-situ}) measurement. Lower detection efficiency in the ToF measurement is due to inelastic molecular collision-induced loss during the additional time of flight compared to the \textit{in-situ} imaging. \textbf{b}, Molecular BEC fraction measured from the focused ToF imaging of the molecular density $n_m$, which shows bimodal distribution at sufficiently low temperature. The inset shows example images of molecules at 44 nK and 302 nK, respectively. The two panels next to the images show line cuts through the image centers, and the blue (red) solid lines represent BEC (thermal) components from a bimodal fit. }
    \label{fig:ExpFig1}
\end{figure}

The procedure to prepare a Cs BEC in the lowest hyperfine ground state at 19.5 G for the quench experiments shown in Fig.~\ref{fig:FigAtm}(a-b) is the same as that in Ref.~\cite{Zhang2023}, where atoms are in a pure optical trap without magnetic field gradient for levitation. The atomic BECs have 23,000 atoms with a BEC fraction of 80\%. We detect the remaining atoms after the quench dynamics by absorption imaging the atoms back at the off-resonant field value 19.5 G. We detect the created Cs$_2$ molecules, in the $g$-wave state $|f=4,m_f=4;\ell=4,m_\ell=2\rangle$, by first blowing away the remaining atoms using the atom imaging light pulse, releasing molecules into a weak horizontal harmonic trap with $\omega_x = \omega_y = 2\pi\times 15$~Hz and wait for a quarter trap period $t_q = 17$~ms. Finally, we image the molecules by jumping the field up to 20.4 G to dissociate them into atoms within 0.1 ms in the optical trap and then image the atoms from the dissociation~\cite{Zhang2021,Zhang2023}. We normalize both the atomic and molecular population by the initial total atom number during the quench dynamics as shown in Fig.~\ref{fig:FigAtm}(a-b). The missing fraction is due to various loss processes. We extract the asymptotic molecular fraction in the quasi-steady state, as presented in Fig.~\ref{fig:3}(a), by averaging data in the time window between 1 ms and 3 ms in the dynamics.

To create pure molecular samples used for the experiments shown in Fig.~\ref{fig:FigAtm}(c), we first make evaporatively cooled ultracold atomic gases at 20.22 G,
where the magnetic field is calibrated \textit{in-situ} by atomic microwave spectroscopy. 
Then we switch to 19.89 G and ramp through the narrow g-wave Feshbach resonance to 19.83 G in 1.5 ms to associate atoms into molecules. After that, we quench the magnetic field to 19.5 G and apply a resonant light pulse to blow away the residual atoms. The resulting molecular temperature and population are characterized and shown in Fig.~\ref{fig:ExpFig1}a, where fewer and colder molecules are created from initial atomic gases at lower temperature and population. When the temperature is low enough, the molecular density after the focused time-of-flight starts to develop a bimodal distribution, from which we do fitting to extract the molecular BEC fraction (see Fig.~\ref{fig:ExpFig1}b). We choose to use molecular BECs at 27 nK with a BEC fraction of 23(1)\% as the initial condition for the experiments shown in Fig.~\ref{fig:FigAtm}(c). After the magnetic field quench and a variable hold time, the atoms from molecule dissociation are imaged \textit{in-situ} for higher detection efficiency.

\section{Quantifying the Resonance Width} \label{app:width}
\vskip2mm

Early experiments on a bosonic Feshbach resonance by Donley et. al. ~\cite{Donley2002,Claussen2002} have focused on coherent oscillations between different Bose condensates below but near resonance.
However, the Feshbach resonance employed, that of $^{85}$Rb atoms at magnetic field $B_0=154.9$G, is very wide. 

Using the many-body classification of resonance width in Ref.~\cite{Ho2012} (see Scheme (B) in their Eq. (4)), we estimate the dimensionless resonance-width parameter to be $x=(k_n r_*)^{-1}\sim 10^{3} \gg 1$ for $^{85}$Rb. For details, see Table~\ref{tab:CsRbK}. 
Here, $k_n= (6\pi^2 n)^{1/3}$ with $n$ the total atomic number density, and $r_*$ is a length scale defined from the experimental resonance width.
As a consequence of the extremely large $x$, the closed channel molecular fraction near unitarity in these wide resonances is negligible~\cite{Kokkelmans2002a}.
And the observed coherent oscillations in Refs.~\cite{Donley2002,Claussen2002} are best interpreted as that
between atomic and ``molecular"-bound states, the latter of which are made up of  open-channel atoms~\cite{Koehler2003,Kokkelmans2002a,MunozdelasHeras2019,Corson2015} and should be contrasted with
the actual closed-channel molecules.

In contrast, the $^{133}$Cs g-wave resonance used in Refs.~\cite{Zhang2021,Zhang2023} is extremely narrow. A simple estimate shows that $x \sim 0.1 \ll 1$,
in agreement with the significant fraction of closed-channel molecules observed near unitarity in the experiments.
The successful observation of molecules in this resonance not only enables us to explicitly study dissociation of molecular superfluids,
but also provides us an opportunity to explore the role of the molecular superfluid component in post-quench dynamics starting with an initial state of
open channel (atomic) superfluid condensate. Theoretically, the inherent narrowness of the resonance requires us to consider a fully two-channel formulation, in order to
treat the dynamics adequately.

In Table~\ref{tab:CsRbK} we present the relevant experimental parameters that we have used to estimate the resonance width for $^{133}$Cs, $^{85}$Rb, and $^{39}$K. 

\section{Derivation of Eq.~\eqref{eq:dynamics} in the Main Text} \label{app:Eq4derivation}

In this section, we give detailed derivations of Eq.~\eqref{eq:dynamics} in the main text. We start with the time-dependent trial wavefunction~\cite{Corson2015}, 
\begin{align}
\vert \Psivar (t) \rangle & = \frac{1}{\mathcal{N}(t)} \nonumber \exp \bigg\{ \sum_{\sigma=1}^2 \Psi_{\sigma 0}(t) \sqrt{V}  a^\dagger_{\sigma \veck=0}   \nonumber \\
& + \sum_{\veck \ne 0}^\prime \sum_{\sigma=1}^2  \chi_{\sigma \veck}(t) \; a^\dagger_{\sigma \veck} a^\dagger_{\sigma -\veck} \bigg\} \vert 0 \rangle, \label{eq:Psivart}
\end{align}
where the prime sign in $\sum_{\veck \ne0}^\prime$ indicates the sum is only over half momentum space such that each $\{\veck,-\veck\}$ pair is counted only once. 
\begin{align}
\mathcal{N}(t) & = \exp( \sum_\sigma |\Psi_{\sigma0}(t)|^2 V/2)
  \prodprime_{ \veck \ne 0} \prod_\sigma (1- |\chi_{\sigma \veck}(t)|^2)^{-1/2}
\end{align}
is the normalization factor. 
In the exponent of Eq.~\eqref{eq:Psivart}, $\Psi_{\sigma 0}$ and $\chi_{\veck}$ are (complex) variational parameters, which are time-dependent for the study of dynamics. 
$\sum_{\veck \ne 0}^\prime = (1/2) V \int^\Lambda d \veck/(2\pi)^3$  with $V$ the volume and $\Lambda$ a cutoff, needed to avoid ultraviolet divergence. 
$|0\rangle$ is the vacuum that is annihilated by all $a_{\sigma\veck}$.

\begin{widetext}
Assuming that the two-channel system, even when it is out of equilibrium, can be always approximated by $ \vert \Psivar(t) \rangle$,
one maps the underlying quantum dynamics, described by the exact Heisenberg equation with the Hamiltonian $\hat{H}$,
to that of a classical system. 
The latter is derived from the action~\cite{Haegeman2011,Shi2018,Kramer2008},
\begin{align} 
 \mathcal{S}[\Psi_{\sigma 0}^*(t), \Psi_{\sigma 0}(t), \chi_{\sigma \veck}^*(t),  \chi_{\sigma \veck}(t)] 
 & =  \int d t  \bigg\{ \langle \Psivar(t) | (i \hbar) \partial_t \Psivar(t) \rangle -  \langle \Psivar(t) \vert \hat{H} \vert \Psivar(t) \rangle   \bigg\}       \label{eq:actionS}  \\
& \equiv \int dt L(\Psi_{\sigma 0}^*(t), \Psi_{\sigma 0}(t), \chi_{\sigma \veck}^*(t),  \chi_{\sigma \veck}(t)). 
\end{align}
Using $\Psivar (t)$ and the Hamiltonian in Eq.~\eqref{eq:Hamiltonian} of the main text, we evaluate the two terms on the right hand side of Eq.~\eqref{eq:actionS} as follows. (For brevity we will suppress all the time dependences in the following.)
\begin{align} 
\langle \Psivar | (i\hbar) \partial_t \Psivar \rangle   & =  \sqrt{V}  \sum_\sigma \big[ ( i \hbar )\frac{d}{dt} \Psi_{\sigma 0} \big]  \langle a^\dagger_{\sigma \veck=0}  \rangle 
  +  \sumprime_{\veck \ne 0} \sum_{\sigma} \big[ ( i \hbar ) \frac{d}{dt} \chi_{\sigma \veck}\big] \langle a^\dagger_{\sigma \veck} a^\dagger_{\sigma -\veck} \rangle  + ( i \hbar ) \frac{d}{dt} \ln \mathcal{N}^{-1}  \\
&   = V \sum_\sigma  \frac{i \hbar}{2} \big(  \Psi_{\sigma 0}^*   \frac{d}{dt} \Psi_{\sigma 0}   -  \Psi_{\sigma 0}  \frac{d}{dt} \Psi_{\sigma 0}^*  \big) 
+  \sumprime_{\veck \ne 0} \sum_\sigma  \frac{  1 } {  1-|\chi_{\sigma \veck}|^2}  \frac{i \hbar}{2} \big( \chi_{\sigma \veck}^* \frac{d}{dt}  \chi_{\sigma \veck}   -   \chi_{\sigma \veck} \frac{d}{dt}  \chi_{\sigma \veck}^*  \big),  \label{eq:PsidtPsi}
\end{align}
where we have introduced the short hand notation, $\langle \cdots \rangle \equiv \langle \Psivar | \cdots | \Psivar \rangle$. 
The other term on the right hand side of Eq.~\eqref{eq:actionS} is given by
\begin{align} 
&  \langle \Psivar | \hat{H} | \Psivar\rangle   = h_{\sigma \veck=0} V  |\Psi_{\sigma 0}|^2 +  \frac{g_\sigma }{2}  V |\Psi_{\sigma 0}|^4   - \alpha V ( (\Psi_{10}^*)^2 \Psi_{20} +c. c.)   \nonumber \\
&  +    \sum_{\veck \ne 0 } \sum_\sigma  ( h_{\sigma \veck} + 2  g_\sigma |\Psi_{\sigma 0}|^2 + g_\sigma n_\sigma)  n_{\sigma \veck }  
  +     \sum_{\veck \ne 0 } \sum_\sigma  \frac{g_\sigma}{2} \big( (\Psi_{\sigma 0}^*)^2   x_{\sigma \veck } + c.c.  \big) + \sum_{\sigma} \frac{g_\sigma}{2} V |x|^2
 -  \alpha \sum_{\veck \ne 0} \big(  \Psi_{20} x^*_{1 \veck}  + c.c.   \big).   \label{eq:PsiHPsi}
\end{align}
\end{widetext}
In arriving at Eqs.~\eqref{eq:PsidtPsi} and \eqref{eq:PsiHPsi} we have used 
\begin{subequations}
\begin{align}
\Psi_{\sigma 0}      &    \equiv   \langle a_{\sigma 0} \rangle/\sqrt{V},   \label{eq:Psi0def} \\
 x_{\sigma \veck}   &   \equiv  \langle a_{\sigma \veck}  a_{\sigma -\veck} \rangle =  \chi_{\sigma \veck} / ( 1-|\chi_{\sigma \veck}|^2),                   \label{eq:xkdef}  \\
n_{\sigma \veck}   &    \equiv \langle a_{\sigma \veck}^\dagger a_{\sigma \veck} \rangle =  |\chi_{\sigma \veck}|^2 / ( 1-|\chi_{\sigma \veck}|^2),   \label{eq:nkdef} \\
 x _\sigma            &     =  V^{-1} \sum_{\veck \ne 0} x_{\sigma \veck} ,  \quad   n_\sigma   =  V^{-1} \sum_{\veck \ne 0} n_{\sigma \veck}.              \label{eq:xxnn}
\end{align}
\end{subequations}
$x_{\sigma \veck}$ is the expectation value of the (Cooper-like) pairing field for atoms ($\sigma=1$) or molecules ($\sigma=2$).

Minimizing $\mathcal{S}$ with respect to $\{\Psi_{\sigma 0}^*,  \chi_{\sigma \veck}^* \}$ leads to the following Euler-Lagrange equations: 
\begin{align}
&  \quad \quad  \frac{\partial L }{\partial \Psi_{\sigma 0}^{*}}  - \frac{d}{dt} \frac{\partial L}{\partial  ( \partial_t \Psi_{\sigma 0}^{*})} =0  \\
&   \Rightarrow 
V (i\hbar)  \frac{d}{dt} \Psi_{\sigma 0} = \frac{\partial}{\partial \Psi_{\sigma 0}^*} \langle \hat{H} \rangle,   \label{eq:EulerL1} \\
& \quad \quad \frac{\partial L }{\partial \chi_{\sigma \veck}^{*}}  - \frac{d}{dt} \frac{\partial L}{\partial   (\partial_t \chi_{\sigma \veck}^{*} ) } =0   \\
 &  \Rightarrow  \frac{1}{(1-|\chi_{\sigma \veck}|^2)^2}  \;  (i\hbar)   \frac{d }{dt} \chi_{\sigma \veck}  =  \frac{\partial}{\partial \chi_{\sigma \veck}^*} \langle \hat{H} \rangle.  \label{eq:EulerL2}
\end{align}
From Eq.~\eqref{eq:EulerL2} and its complex conjugate we then derive
\begin{align}
i \hbar \frac{d}{dt} x_{\sigma \veck }   & =  \frac{i \hbar}{(1-|\chi_{\sigma \veck}|^2)^2} \big(  \frac{d \chi_{\sigma \veck} }{dt}   +  \chi_{\sigma \veck}^2  \frac{d \chi^*_{\sigma \veck}}{dt} \big) \\
& = \frac{\partial}{\partial \chi_{\sigma \veck}^*} \langle \hat{H} \rangle   -  \chi_{\sigma \veck}^2 \frac{\partial}{\partial \chi_{\sigma \veck}} \langle \hat{H} \rangle,  \label{eq:EulerL2v2}
\end{align}
where we have used Eq.~\eqref{eq:xkdef}. 
Substituting the expression of $\langle \hat{H} \rangle$ from Eq.~\eqref{eq:PsiHPsi} into Eqs.~\eqref{eq:EulerL1} and ~\eqref{eq:EulerL2v2} leads to
\begin{widetext}
\begin{subequations} \label{eq:dynamics2}
\begin{align}
i \hbar  \frac{d}{dt} \Psi_{1 0}     & =  (  h_{1\veck=0}  + g_1 |\Psi_{10}|^2  + 2 g_1 n_1  ) \Psi_{10} + g_1 \Psi_{10}^* x_1 - 2 \alpha  \Psi_{10}^* \Psi_{20}, \\
i \hbar  \frac{d}{dt} \Psi_{2 0}      & = (  h_{2 \veck=0}  + g_2 |\Psi_{20}|^2  + 2 g_2 n_2 ) \Psi_{20} + g_2 \Psi_{20}^* x_2 - \alpha  ( x_1 + \Psi_{10}^2 ), \\
i \hbar \frac{d}{dt} x_{1 \veck}   &  =  2  \big[  h_{1\veck} + 2 g_1 ( |\Psi_{10}|^2 + n_1)   \big] x_{1\veck}   +  \big[g_1 (x_1 + \Psi_{10}^2)  - 2 \alpha \Psi_{20} \big] (2 n_{1\veck}+1) ,  \\
i  \hbar \frac{d}{dt} x_{2 \veck}   &  = 2 \big[  h_{2\veck} + 2 g_2  ( |\Psi_{20}|^2 + n_2)     \big]  x_{2\veck} +   g_2 (x_{2}  +  \Psi_{20}^2) (2 n_{2\veck}+1). 
\end{align}
\end{subequations}
\end{widetext}
We emphasize that in evaluating the partial derivative, $ \partial \langle \hat{H} \rangle / \partial \chi_{\sigma \veck}^*$, to obtain the last two equations, we have to include contributions from terms in $\langle \hat{H} \rangle$ (Eq.~\eqref{eq:PsiHPsi})  both at $\veck$ and $-\veck$, as each $\{\veck, -\veck\}$ pair shares the same variational parameter $\chi_{\sigma \veck}$ in the exponent of our variational wavefunction (see Eq.~\eqref{eq:Psivart}). 
Otherwise, the $d x_{\sigma \veck}/dt$ obtained will differ from the above expressions by a factor of $2$.

\subsection{An alternative derivation}
In this subsection we sketch an alternative derivation for Eq.~\eqref{eq:dynamics2}, which shows more explicitly in what sense the quantum dynamics can be mapped to the classical-dynamics described by the action $\mathcal{S}$ in Eq.~\eqref{eq:actionS}. It may also help us to better understand when the classical equations derived from $\mathcal{S}$ will become inadequate in future applications, although such a potential breakdown is not of the major concern to our current paper. 

In this alternative approach, we start with the exact Heisenberg equation for a generic operator $ \hat{O}(t) \equiv e^{i \hat{H} t/\hbar} \hat{O} e^{- i \hat{H} t/\hbar}$,
\begin{align}
\frac{d\hat{O}(t)}{dt}  & = \frac{ i}{\hbar} [\hat{H},\hat{O}(t)]. \label{eq:Heisenberg}
\end{align}
Within our current variational wavefunction scheme, $\hat{O}$ can be either $a_{\sigma \veck=0}$ or $a_{\sigma \veck} a_{\sigma -\veck}$. 
Next, we make the following central approximation,
\begin{align}
\frac{d}{dt}  \langle  \hat{O}(t)  \rangle \approx \langle  \frac{d\hat{O}(t)}{dt}  \rangle = \frac{ i}{\hbar}  \langle  [\hat{H},\hat{O}(t)] \rangle.  \label{eq:Heisenberg2}
\end{align}
From this equation, we then derive Eq.~\eqref{eq:dynamics2} as an approximation to the exact Heisenberg quantum dynamics.

First, consider $\hat{O}=a_{\sigma \veck=0}$. From Eq.~\eqref{eq:Heisenberg2} one can show that 
\begin{align}
V (i\hbar) \frac{d}{dt} \Psi_{\sigma 0}  & = i \hbar \sqrt{V} \frac{d}{dt} \langle  a_{\sigma \veck=0}  \rangle \approx  \sqrt{V} \langle   [\hat{O} ,\hat{H} ] \rangle  \nonumber  \\
& =  \frac{\partial}{\partial \Psi_{\sigma 0}^*} \langle  \hat{H}  \rangle.
\end{align}
Apart from the approximate sign, this equation is identical to Eq.~\eqref{eq:EulerL1}. 
Similarly, it follows that for the Cooper-like pairing field $\hat{O}= a_{\sigma \veck} a_{\sigma -\veck}$, 
\begin{align}
i \hbar \frac{d}{dt} x_{\veck} & = i \hbar \frac{d}{dt} \langle  a_{\sigma \veck} a_{\sigma -\veck}  \rangle  \approx \langle  [\hat{O} , \hat{H} ]  \rangle  \nonumber \\
&  =   \frac{\partial}{\partial \chi_{\sigma \veck}^*} \langle  \hat{H} \rangle -   \chi_{\sigma \veck}^2  \frac{\partial}{\partial \chi_{\sigma \veck}}  \langle  \hat{H} \rangle,
\end{align}
which is essentially identical to Eq.~\eqref{eq:EulerL2}. The remaining derivations leading to Eqs.~\eqref{eq:dynamics2} are the same as in the previous section.

\section{Regularization and Renormalization}
\label{app:regularization}

Because we have used contact interactions in the Hamiltonian Eq.~\eqref{eq:Hamiltonian} in the main text, solving Eq.~\eqref{eq:dynamics2} requires a proper regularization to avoid
ultraviolet divergences in integrals over $\veck$. The regularizations can be determined by matching the equilibrium version of Eq.~\eqref{eq:dynamics2} with the corresponding Lippman-Schwinger equation in the two-body scattering limit as done in Ref.~\cite{Kokkelmans2002}. 

For the open channel atoms, a correct renormalization condition, that is compatible with the definition of $\hat{H}$ in Eq.~\eqref{eq:Hamiltonian} of the main text, is given as follows~\cite{Kokkelmans2002a}, 
\begin{subequations} \label{eq:regularization}
\begin{align}
g_1   & = \bar{g}_1 \Gamma, \\
\alpha & = \bar{\alpha} \Gamma / \sqrt{2},\\
 \nu  & = \bar{\nu}  + \sqrt{2} \beta \alpha \bar{\alpha},
\end{align}
\end{subequations}
with
\begin{subequations}
\begin{align}
\bar{g}_1 & = \frac{4\pi \hbar^2 \asbg}{m_1}, \\
\beta & =  \frac{m_1 \Lambda}{2\pi^2 \hbar^2 }.  \\
\Gamma & =\frac{1}{1-\beta \bar{g}_1}, \\
\bar{\alpha}^2 & =\bar{g}_1 \Delta \mu_m \Delta B, \\
 \bar{\nu}   & = \Delta \mu_m (B- B_0). 
\end{align}
\end{subequations}
In these equations, quantities denoted with a bar atop represent the renormalized (or physical) ones that are directly related to experimental observables, while those without the bar are bare ones whose
value depends on the cutoff $\Lambda$.  $\asbg$ is the atom-atom background scattering length. $B$ is the applied external magnetic field in experiments, and $B_0$ corresponds to the resonance point where the atom-atom scattering length diverges. 
 $\Delta B$ is the resonance width measured in magnetic fields, and  $\Delta \mu_m$ is the magnetic moment difference between a pair of atoms in the open channel and a molecule in the closed channel. 

One can also derive the regularization and renormalization relations in Eq.~\eqref{eq:regularization} directly from Eq.~\eqref{eq:dynamics2} by considering the zero-density limit of the latter. 
In this limit, we ignore the dynamics of $\Psi_{20}$ and $x_{1\veck}$ in Eq.~\eqref{eq:dynamics2}, integrate them out, subsum their effects into the equation for $i \hbar d\Psi_{10}/dt$, and cast the obtained results into a form of Gross-Pitaevskii equation for $\Psi_{10}$,
with the following effective atom-atom interaction parameter
\begin{align}
g_{1,\mathrm{eff}} & = \frac{g_1}{1+ g_1 \beta} -  \frac{2 \alpha^2/(1+ g_1 \beta)^2}{\nu - 2  \beta  \frac{\alpha^2}{1+ g_1 \beta}}. \label{eq:asdef1}
\end{align}
This $g_{1,\mathrm{eff}}$ is identified with $4\pi \hbar^2 a_s/m_1$, where $a_s$ is the $\nu$-dependent atom-atom scattering length. 
Comparing this result with the definition of $a_s$ in terms of physical observables,
\begin{align}
 a_s &  \equiv \asbg  -   \frac{m_1} {4\pi \hbar^2 }   \frac{\bar{\alpha}^2}{\bar{\nu}} = \asbg  \bigg( 1 - \frac{\Delta B}{B - B_0} \bigg),  \label{eq:asdef2}
\end{align}
we immediately see that Eq.~\eqref{eq:regularization} is a correct renormalization condition. 
In Eq.~\eqref{eq:asdef2}, $\bar{\alpha}$ measures the Feshbach resonance width in units of energy, and $\bar{\nu}=\Delta \mu_m (B-B_0)$ is the detuning measured in energy.

For the closed channel molecules, the proper regularization that connects the bare interaction parameter $g_2$ to the molecule-molecule background scattering length $\ambg$ is given by the following
Lippman-Schwinger equation,
\begin{align}
\frac{m_2}{4\pi \hbar^2 \ambg} & = \frac{1}{g_2} + \int^\Lambda \frac{d \veck}{(2\pi)^3} \frac{1}{ \hbar^2 \veck^2/m_2}, \label{eq:g2reg}
\end{align}
where $m_2 =2 m_1$ is the molecule mass and $\ambg$ is the molecule-molecule background scattering length. In principle, the cutoff $\Lambda$ here can be different from the one used in Eqs.~\eqref{eq:regularization}. Here, we take them to be the same. 

From the experimental values of $\{ \asbg, \Delta \mu_m, \Delta B \} $ from Table~\ref{tab:CsRbK} and $\ambg=220 a_B$, taken from Refs.~\cite{Zhang2021,Zhang2023}, we determine the bare interaction parameters $\{\alpha, g_1, g_2\}$ in our Hamiltonian for a chosen cutoff $\Lambda$, using the renormalization conditions in Eqs.~\eqref{eq:regularization} and ~\eqref{eq:g2reg}. 
In our numerics we leave the cutoff $\Lambda$ as a relatively free parameter, which is adjusted such that the resulting results are in reasonable agreements with the experiments
at comparable detuning. In Table.~\ref{tab:parameters} we list the parameters $\{g_1, g_2, \alpha, \Lambda\}$ that we have used in our simulations.


\section{Understanding the Pairing Contributions}
\label{app:pair}

In this section we give a more extensive discussion of the pairing
contributions which appear after a quench of an atomic condensate as the quenched detuning is varied towards resonance and even beyond.
It is shown here that this introduction of the pairs which occurs during
the transient stage, essentially instigates the subsequent dynamics.

There seem to be two schools of thought on the atom-molecule dynamics. In the first
of these all dynamical processes and oscillations are associated with the condensates
only~\cite{Vardi2001} (although fluctuation effects can also
be contemplated), whereas in the second~\cite{Kokkelmans2002a,Holland2001a}
pairing contributions are important, although they have not been treated before in the
presence of a substantial fraction of closed-channel superfluid molecules.
It should be clear here that our approach is to be distinguished from the
condensates-only scheme. Notably in Ref.~\cite{Zhang2023} such an approach
was taken but in the context of an extended 3-body interaction term. One
can, in fact, make a case that the 3-body Hamiltonian introduced in Ref.~\cite{Zhang2023} will be in some sense an effective interaction between condensed atom and molecules,
mediated by pairs through higher order (in Feshbach coupling) contributions of the latter.

\begin{figure}[h]
\centering\includegraphics[width=\linewidth]{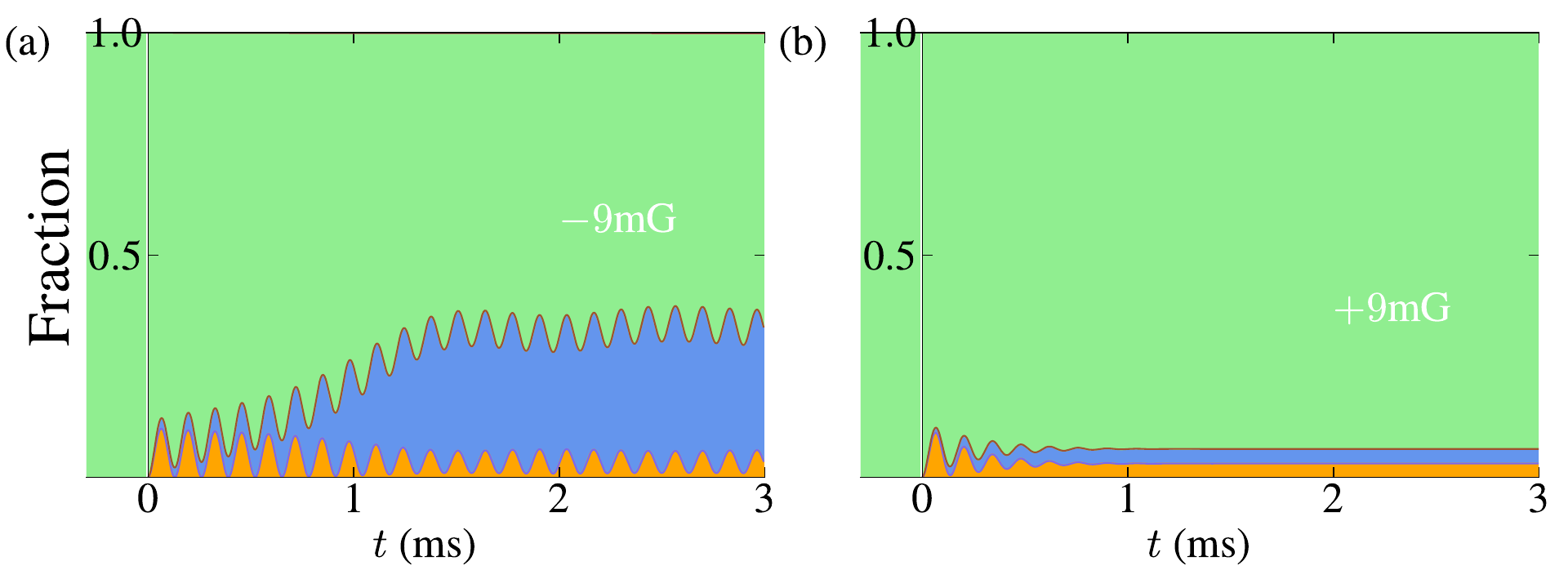}
\caption{Contrast between the quench dynamics at large negative (panel (a)) and positive ((b)) detuning. Shaded green, blue, and orange regimes represent the atomic condensate, non-condensed pair, and molecular condensate fraction, respectively. Indicated in white are the quenched detuning $\bar{\nu} /\Delta \mu_m =  B -B_0$ (in mG). }
 \label{fig:pair1}
\end{figure}

We begin with Fig.~\ref{fig:pair1}(a) which addresses a sweep from an atomic 
condensate to rather further to the molecular side of resonance than in Fig.~\ref{fig:4}(a) in the main text. It is worth concentrating
on the detailed time dependence as this shows that in the early stages of the evolution the greatest change is associated with the creation of a molecular
condensate. But shortly thereafter the pairing contribution begins to grow.
In this case an overall envelope shows that the pairing is growing at the
expense of the atomic condensate and this is expected because this sweep is deeper on the
molecular side so that the initial atomic condensate is less stable.  
After a transient, the molecular condensate is frozen and rather time
independent except for small oscillations. In addition, there is a three-way
coupled oscillation between the atomic and the molecular condensates and the pairs.

If we compare Fig.~\ref{fig:pair1}(a) with ~\ref{fig:pair1}(b) 
where the final state of the system is on the atomic side of resonance, it is
clear that the molecular condensate and the pairing terms are in this new figure much
reduced in magnitude.
In Fig.~\ref{fig:pair1}(b), one sees that the atomic condensate is not as driven to decay, since it is not as unstable
as in the previous case.
Hence we see fewer pairs.
Here, too, one sees after a transient that there is a three-way coupled oscillation.

\begin{figure}[h]
\centering\includegraphics[width=\linewidth]
{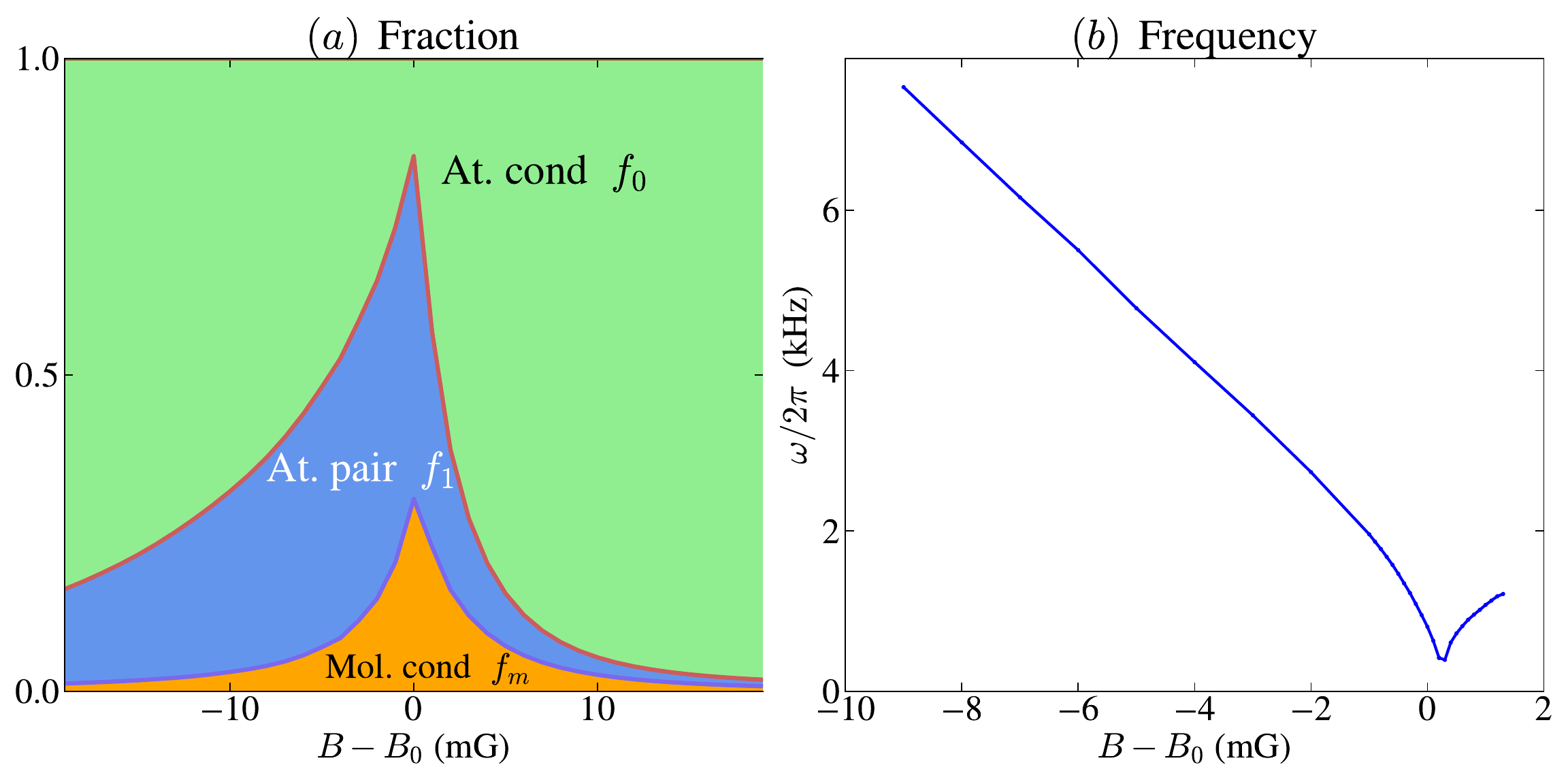}
\caption{ Panel (a) shows the time averaged weight of each of the three components (atomic condensate fraction $f_0=|\Psi_{10}|^2/n$,  non-condensed atom-pair fraction $f_1=n_1/n$,
and molecule fraction $f_m=2 |\Psi_{20}|^2/n$) as a function of the quenched detuning $\bar{\nu}/\Delta \mu_m = B -B_0$. The results are obtained for the steady state reached after a quench as in Fig. ~\ref{fig:FigAtm}(a,b) in the main text. It relates to Fig. ~\ref{fig:3}(a) in the
main text by showing the quantities that were not plotted in Fig. ~\ref{fig:3}(a), namely $f_0$ and $f_1$; the results here could serve as a good basis for predictions to
be addressed experimentally in future. (b) This figure plots the steady-state oscillation frequency $\omega$
as a function of the quenched detuning $B-B_0$ for fixed particle number density $n=2.9\times 10^{13}$cm$^{-3}$. 
}
 \label{fig:pair2}
\end{figure}

We next turn to the component contributions for more general situations where
the final state detuning is varied continuously. This is plotted in Fig.~\ref{fig:pair2}(a).
This figure can be compared to Fig.~\ref{fig:3}(a) in the main text. What is most
striking here is that 
while the molecular boson contributions are reasonably symmetric
around resonance, the pair contribution is more
significant on the molecular side, as already seen in Fig.~\ref{fig:pair1}.
Indeed, we have argued in the text for such an asymmetry based on
energy conservation issues. When the molecular level is far below
the atomic level the creation of molecules must be compensated by
introducing higher energy states, in this case pairs.

In addition to this asymmetry what is rather interesting here is that
there is a re-stabilization of the atomic condensate deep into the
molecular side of resonance. This is rather similar to what one would observe in a simple two-level Rabi oscillation.

For completeness, we also show in Fig.~\ref{fig:pair2}(b) the steady-state oscillation frequency vs detuning $B-B_0$.
There is a clear V-shape with a minimum of frequency at $B-B_0$ very
close to zero, corresponding to the $2$-body resonance, but more
precisely at $B-B_0\approx 0.25$mG. Here the plot terminates at
$B-B_0 = 2$mG, because the oscillations above $2$mG are completely damped.

In summary, given that there is a dichotomy between pairing contributions
and condensate-only contributions (but which go beyond the simple
two-body Feshbach coupling), it will be important in the future to obtain more direct
experimental evidence for or against these non-condensed pair effects.
Similarly, for future theory it may be important to
include direct pair-wise inter-condensate correlations.

\section{More details on theory-experiment comparison: dependence of the oscillation frequency 
on the particle density}
\label{app:comparison}

\begin{figure}[htp]
\begin{center}
\includegraphics[width=0.8\linewidth]{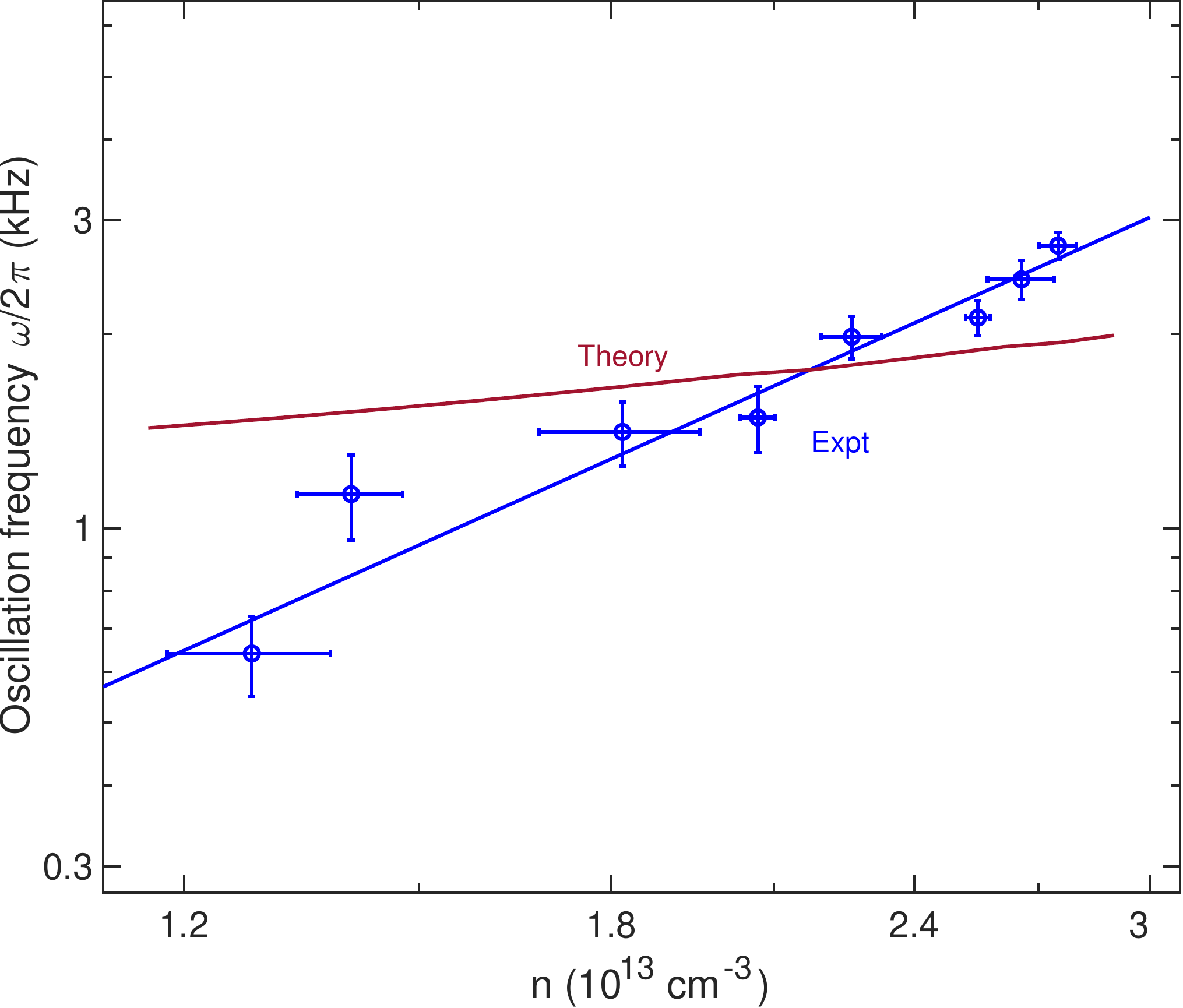}
\caption{Density ($n$) denpendence of the oscillation frequency ($\omega$) near unitarity. 
The experimental data are the same as in Fig.~\ref{fig:3}(b) of the main text.
The theoretical curve, in magenta, is taken at $B-B_0=-1$ mG, 
which should be contrasted with the theoretical curve in Fig.~\ref{fig:3}(b) of the main text, 
which is for $B-B_0=0$ mG. 
The theoretical curve at $B-B_0=-1$ mG here roughly follows $\omega \propto n^{0.6}$ within the density range plotted. 
The comparison here is to show that if one takes into account the fact that the experimental data is collected for $B-B_0=-1$ mG, 
a much better agreement between the magnitude of the theoretical and experimental oscillation frequencies can be obtained. }
\label{fig:Fig4anew}
\end{center}
\end{figure}

It is important to point out that the experimental data in Fig.~\ref{fig:3}(b) are collected for $B-B_0=-1$ mG, which is
not strictly at unitarity where the theory was addressed. 
There is some experimental uncertainty ($\sim$2 mG) in the measured B field, which mainly comes from environmentally-caused stray fields (of about $14$ mG). We suppress the stray fields by a servo loop to the level of 2 mG.

Given this uncertainty, if we use our theoretical result at $B-B_0=-1$ mG to compare with the experiment,
as shown in Fig.~\ref{fig:Fig4anew}, we see that the oscillation frequency magnitude is actually in rather good agreement with the experimental data.

Importantly, in the context of
Fig.~\ref{fig:Fig4anew},
while there is a discrepancy in the power-law exponent between theory and experiment,
we argue that this does not mean
there is a contradiction between the theoretical model description used in the current paper and the three-body recombination mechanism advocated in Ref. \cite{Zhang2023}.
Even though the microscopic two-channel Hamiltonian we started with only contains Feshbach coupling ($\alpha$) and pair-wise density-density interactions ($g_1$ and $g_2$), it can induce three-body recombination through scattering processes that are higher order than linear in $\alpha$. This should not be surprising since the two-channel Hamiltonian (with point contact interactions) has been used in the literature to discuss three-body recombination and related Efimov physics. See Ref.~\cite{Bedaque2000} for example. The two- and three-body model Hamiltonians used in Ref. \cite{Zhang2023} should be understood as the \textit{full two- and three-body scattering amplitudes} between atom and molecules derived from an infinite sum of microscopic scattering process resulting from the two-channel microscopic Hamiltonian.

\section{Possible causes of the discrepancy between the theory and experiment in the minimal oscillation frequency}

The comparison between our theory and experiment is not perfect. In particular, there is a discrepancy in the minimal oscillation frequency 
as a function of detuning between the experimental results and theory (see Fig.~\ref{fig:pair2} and Fig. 3(d) of Ref.~\cite{Zhang2023}). 
One may speculate that some of the following points, which are largely ignored in the theoretical literature as well as in the current theoretical treatment, have contributed to this discrepancy:

\begin{enumerate}
\item In our theoretical modeling we have ignored a possible inter-channel density-density interaction term,
$g_{12} \sum_{\veck_1,\veck_2,\veck_3} a^\dagger_{1,\veck_1} a_{1,\veck_2} a^\dagger_{2,\veck_3} a_{2, \veck_1+\veck_3-\veck_2}$, which will make additional contributions to
the dynamic equations of both atom and molecule condensates, $d\Psi_{10}/dt$ and $d \Psi_{20}/dt$, if we assume that the $g_{12}$ effect is elastic. 
This additional term depends on the amplitudes of both atom and molecule condensates, $|\Psi_{10}|$ and $|\Psi_{20}|$. 
Given that this term is off-diagonal in the subspace spanned by the atom and molecule condensate energy levels, 
it behaves very much like the inter-level coupling term in a two-level Rabi oscillation problem; therefore, one expects that including this term will lead to a larger minimal oscillation frequency. 
The existence of this contribution, which is proportional to $|\Psi_{10}|$ and $|\Psi_{20}|$,
is also consistent with the observation that the minimal oscillation frequency in Fig. 3(d) of Ref.~\cite{Zhang2023} increases with the initial atom BEC fraction. 
\item Another simplification which we make and which is widespread in the literature is to drop
correlations, such as $\langle a_{1,\veck} a_{2,-\veck}\rangle$, in our many-body trial wavefunction. Including these additional inter-channel correlations, which increases the complexity
significantly, can also affect the minimal oscillation frequency.
\item Lastly, in our theoretical modeling we have ignored various possible loss processes due to atom-atom and atom-molecule inelastic collision. 
\end{enumerate}


%

\end{document}